\documentclass[runningheads]{article}

\usepackage{graphicx}
\graphicspath{ {images/} }
\usepackage[utf8]{inputenc} 
\usepackage[T1]{fontenc}    
\usepackage{hyperref}       
\usepackage{url}            
\usepackage{booktabs}       
\usepackage{amsfonts}       
\usepackage{nicefrac}       
\usepackage{microtype}      
\usepackage{bm}
\usepackage{color}
\usepackage{wrapfig}
\usepackage{algorithm2e}
\usepackage{tcolorbox}
\usepackage{color}
\usepackage{url}
\usepackage{tikz,pgfplots}
\usepackage{pgf}            
\usepackage{multirow}

\usepackage{algorithmic}\usepackage{url}
\usepackage{amsmath}
\newcommand{\R}{\mathbb{R}}
\newcommand{\V}{\mathcal{V}}

\begin{document}

\title{Quels corpus d'entraînement pour l'expansion de requêtes par plongement de mots : application à la recherche de microblogs culturels}

\author{Philippe Mulhem, Lorraine Goeuriot, Massih-Reza Amini, Nayanika Dogra\\
Université Grenoble Alpes, CNRS, Inria, Grenoble INP\\
LIG/IMAG, 700 av. centrale\\
38401 Saint-Martin d'Hères, France\\
\url{FirstName.Lastname@univ-grenoble-alpes.fr}
}
\date{}

\maketitle

\begin{abstract}
Cet article décrit un cadre expérimental et des résultats obtenus pour la recherche de microblogs. Nous étudions la corrélation entre la source d'apprentissage des plongements de mots et l'apport de ces derniers à un système de recherche d'information. Nous étudions en particulier son utilisation pour étendre des requêtes sur des tweets culturels sur le corpus CLEF CMC 2016. Nos résultats montrent que l'utilisation de corpus spécifiques (au niveau sujet ou bien sujet+type de document) ne fournit pas forcément de meilleurs résultats.
\end{abstract}

\section{Introduction}

Le principe général de la recherche d'information est de trouver les documents les plus pertinents répondant à la requête d'un utilisateur. Pour cela, un utilisateur pose une requête qui est traitée par un système de recherche d'information. En réponse, ce système fournit une liste de documents triés par valeurs décroissantes de scores de pertinence calculés par le système.
Dans le cas où les intersections entre les termes des requêtes et des documents sont faibles, il est nécessaire d'intégrer au processus de recherche des similarités entre les mots, afin de diminuer le silence dans les réponses et d'améliorer la pertinence des résultats. Cela est le cas par exemple pour la recherche de microblogs (publications courtes telles que les tweets). Nous proposons d'utiliser des similarités entre mots appris par des plongements de mots, et de les utiliser en étendant les requêtes des utilisateurs. Nous nous situons dans un cadre spécifique : la recherche de microblogs culturels en langue française, et plus particulièrement des tweets émis durant des festival musicaux. La nature des données constitue un défi en soi pour la recherche d'information : les documents sont très courts, et véhiculent un contenu contenant de nombreuses références (à des documents plus longs, à des utilisateurs, à des thématiques - hashtags, etc.). Les requêtes et les documents étant, par nature, très courts, il est nécessaire de recourir à des méthodes permettant d'en étendre le contenu. 
Les corpus d'apprentissage considérés diffèrent du point de vue sujet (général ou relatif à la musique) ainsi que du point de vue du type de texte (général vs thématique, document web vs microblog).

Dans cet article, nous exploitons des méthodes "classiques" d'expansion de requête à l'aide de plongements de mots, en étudiant l'impact d'un paramètre très important de l'apprentissage des plongements, le {\it corpus d'apprentissage}, sur la qualité des résultats de la recherche. Les approches d'expansion utilisées sont tirées de (Almasri et al. 2017, Almasri 2017). Ces approches ont le mérite d'être assez simples : elles exploitent directement les espaces de plongement. Elles peuvent donc nous permettre de mesurer l'apport des plongements de manière directe. Nos résultats montrent qu'il est préférable d'utiliser, dans notre cadre, des plongements appris sur le corpus utilisé par le système de RI (corpus de microblogs sur le sujet spécifique des festivals de musique), même si ce corpus n'est pas très volumineux.

Cet article suit le plan suivant. Nous commençons par décrire des travaux de l'état de l'art relatif à la recherche d'information et à l'utilisation des plongements de mots en partie~\ref{sec:etatdelart}. Nous décrivons ensuite les approches que nous étudions pour l'expansion de requêtes avec des plongements en section~\ref{sec:exprequ}. Nos expérimentations et les résultats obtenus sont rapportés en partie~\ref{sec:expes}, avant de conclure.

\section{État de l'art}
\label{sec:etatdelart}

\subsection{Recherche d'information sur des microblogs}\label{ss:RImicroblogs}
La recherche d’information moderne se base sur de nombreux médias visant à aider les gens à trouver des informations dans des contextes de plus en plus divers. Parmi ces nouveaux médias, il y a ceux qu'on appelle des microblogs. Un microblog est une trame de textes écrits par un auteur au fil du temps. Il comprend de nombreuses mises à jour très brèves qui sont présentées aux lecteurs dans l’ordre chronologique inverse.  Bien que le microblogging soit de plus en plus populaire, les méthodes d'accès aux données de microblog sont encore nouvelles. Parmi ces plateformes, Twitter est celle qui est devenue la plus populaire. 

Les microblogs sont des documents spécifiques pour différentes raisons~\cite{JabeurTB12,Gimpel2011,twdatadict2018} :
\begin{itemize}
    \item ils sont courts (quelques centaines de caractères);
    \item ils utilisent des mots-clés spécifiques (comme des {\it hashtags} dans Twitter) qui permettent de distinguer de manière explicite des sujets dans le microblog;
    \item ils peuvent contenir des liens pointant sur des informations supplémentaires (images, etc.) extérieures au microblog;
    \item ils utilisent des syntaxes et des vocabulaires spécifiques, comme des abréviations, des émoticons, etc.;
    \item ils sont souvent utilisés pour émettre des avis, des opinions~\cite{Mohammad2017}. Ils sont donc fortement subjectifs.
\end{itemize}

Le recherche de tels documents rebat les cartes de la recherche d'information classique qui repose sur l'intersection des termes en requêtes et documents. Dès lors, il est nécessaire de se poser des questions pour tenter d'apporter des réponses à ce cadre spécifique. Une idée assez simple est de rester sur un système de recherche d'information classique, et d'utiliser des ressources externes quelconques pour améliorer l'intersection entre les termes des requêtes et des documents en étendant artificiellement le contenu. Ces ressources externes ont en fait comme objectif de réaliser une expansion des termes des documents ou des requêtes (ou des deux), afin de diminuer un éventuel silence dans les réponses.

\subsection{Représentations de mots}
Les systèmes de traitement automatique du langage naturel (TALN) traitent traditionnellement des mots comme des index atomiques discrets \cite{Salton68}. Ces encodages sont arbitraires et ne fournissent aucune information utile pour le système en ce qui concerne les relations de synonymie ou de polysémie qui peuvent exister entre les différents index de mots et qui sont nécessaires pour comprendre le sens des textes courts comme ceux des micro-blogs.

De nombreux travaux ont montré que la représentation distribuée de mots dans un espace vectoriel continu permet aux systèmes d'atteindre une meilleure performance pour différentes  tâches du TALN \cite{Amini07, Kim08, Pessiot10, Balikas16}, ceci généralement parce qu'elle leur permet de regrouper les mots similaires et de réduire la variabilité dans les vocabulaires des collections associées. 

Il existe deux types d'approches pour apprendre une telle représentation distribuée : les modèles latents à base de comptages statistiques, comme le LDA (\textit{Latent Dirichlet Allocation}) \cite{Ble03} ou le PLSA (\textit{Probabilistic Latent Semantic Analysis}) \cite{hofmann_sigir99}, et les méthodes prédictives à base de réseaux de neurones. Ces deux types sont basés sur l'hypothèse distributionnelle stipulant que les mots apparaissant dans les mêmes contextes avec les mêmes fréquences sont similaires.  Le contexte est généralement une phrase ou un paragraphe.

\subsubsection{Modèles latents}
Les méthodes d'indexation usuelles reposent sur des termes simples, considérés comme indépendants les uns des autres. Toutefois, il existe bien souvent un certain décalage entre les termes utilisés par un auteur dans un microblog et ceux employés par une personne posant une requête. Ce décalage se traduit par le fait que différents termes peuvent être utilisés pour désigner le même concept, ou qu'un même terme désigne  différents concepts suivant son contexte d'usage. La présence de synonymes (ou plus généralement de mots de sens proches) conduit bien souvent à des taux de rappel plus faibles, car des documents pertinents ne sont plus appariés avec la requête. En revanche, le phénomène de polysémie nuit à la précision car des documents non pertinents, portant en fait sur un tout autre sujet, sont appariés avec la requête.
  
De nombreuses propositions ont été avancées pour rendre compte au mieux de ces deux phénomènes, soit au niveau de l'indexation, soit au niveau de l'appariement entre documents et requêtes. Beaucoup de ces propositions reposent sur l'exploitation de ressources linguistiques explicites, telles que dictionnaires, listes de synonymes, thésaurus ou ontologies. De telles ressources permettent d'une part de choisir une forme normalisée pour les différents synonymes d'un même concept (ou plus précisément pour les différents synonymes d'un même concept présents dans les ressources à disposition) et d'autre part, lorsqu'elles sont couplées à des systèmes de désambiguïsation sémantique, de différencier un même terme suivant le contexte dans lequel il est utilisé. Ce genre d'approches présente toutefois deux désavantages majeurs:

\begin{itemize}
\item Tout d'abord, le développement de telles ressources est extrêmement coûteux et l'on est loin de disposer de thésaurus pour tous les domaines et toutes les langues.
\item Ensuite, un langage, fut-il de spécialité, est loin d'être figé. Il emprunte sans cesse à d'autres langues et crée de nouveaux termes pour désigner les nouveaux concepts apparaissant avec le développement technologique et scientifique et avec les nouveaux points de vue que l'on peut porter sur des objets anciens. De plus, les utilisations des termes en corpus varient d'un individu à l'autre et ne correspondent pas toujours aux usages définis dans des ressources lexicographiques.
\end{itemize}

Les ressources dont on peut disposer à un instant donné ne seront donc que partielles et ne fournissent pas une solution complète aux problèmes de synonymie et polysémie. Beaucoup de chercheurs se sont alors tournés vers d'autres approches pour construire, de façon automatique, des représentations qui rendent compte de certaines relations entre mots (synonymie) et qui permettent à un même mot d'appartenir à des représentations différentes (polysémie). 

\paragraph{\textbf{Analyse Sémantique Latente}.} Cette analyse consiste à effectuer une décomposition en valeurs singulières de la matrice termes-documents. Une telle décomposition fournit en effet un espace latent\footnote{Le terme latent signifie ici que cet espace n'est pas directement observable mais est sous-jacent à la collection étudiée.} dans lequel documents et termes sont représentés simultanément et dans lequel chaque axe peut être vu comme un concept sous-jacent à la collection. Plus précisément, l'analyse sémantique latente fournit la projection des documents et des termes de la collection dans cet espace sous-jacent, la projection des termes étant utilisable pour construire la représentation de nouveaux documents (comme les requêtes). L'analyse sémantique latente souffre néanmoins d'une limitation majeure qui est celle de la difficulté d'interpréter les matrices obtenues. En effet, le fait que ces matrices contiennent des éléments négatifs ne permet pas vraiment de représenter les facteurs latents obtenus : que signifie un poids négatif d'un terme sur un axe latent? Un poids de 0 fait conclure qu'un terme ne participe pas à la définition de l'axe, mais un poids négatif est beaucoup plus difficile à interpréter.

\paragraph{\textbf{Analyse Sémantique Latente Probabiliste}.} Le cadre probabiliste de l'analyse sémantique latente (\textit{PLSA}) a été proposé pour la première fois, dans~\cite{hofmann_tr98}, puis dans \cite{hofmann_sigir99} pour son application à l'indexation. Il s'appuie sur une vision générative de la co-occurrence entre un document et un terme, et il associe une variable de classe non observée (ou latente)  à chaque occurrence de terme. L'estimation des paramètres du modèle, est réalisée suivant le principe de maximisation de la vraisemblance, par l'intermédiaire de l'algorithme EM \cite{Dempster77}.

\paragraph{\textbf{Latent Dirichlet Allocation}. }
Le modèle PLSA ne permet pas de générer de nouvelles données, ainsi pour extraire les thèmes latents d'un nouveau document dans la collection, il faudrait à nouveau ré-estimer les probabilités des termes sachant les thèmes, ce qui devient vite handicapant avec des collections en constant évolution comme celles des micro-blogs. Afin de remédier à ce défaut génératif du modèle PLSA, tout en conservant ses bonnes propriétés d'affectation souple, \cite{Ble03} ont introduit un nouveau modèle latent, dénommé LDA. Dans ce modèle, chaque
terme d'un document est généré à partir d'un mélange de multinomiales sur un
ensemble de thèmes latents. L'ensemble des thèmes suit une loi multinomiale de paramètres
 qui suivent eux-mêmes une loi de Dirichlet. La distribution de
Dirichlet est une généralisation multivariée de la distribution bêta. 

\subsubsection{Approches à base de connaissances structurés}
Il est aussi possible de considérer l'utilisation de bases de connaissances structurées existantes pour retrouver des tweets. \cite{QiangFLY15} propose d'utiliser la base {\it Freebase} pour extraire des termes qui étendent la requête. Même si la base de connaissance utilisée est grande, le problème de correspondance entre les termes de la base et ceux des documents est dans ces cas difficile à régler. La solution prise par~\cite{QiangFLY15} (qui reprend les idées de~\cite{KimYC12}) est d'utiliser du pseudo bouclage de pertinence sur les premiers documents renvoyés. La manière de mettre en correspondance les termes des bases de connaissances et des documents repose sur la qualité des bases de connaissance et leur structuration, et ces approches sont peu robustes à des évolutions de ces bases.
  
\subsubsection{Réseaux profonds}
 Les modèles prédictifs sont inspirés du travail séminal de \cite{Rumelhart88}.  Le but de ces modèles est de trouver une fonction d'association qui prend un mot en entrée et qui le projette dans un espace vectoriel de dimension réduite.  Les modèles prédictifs neuronaux sont généralement des modèles récursifs qui, en partant d'une structure syntaxique d'un contexte, associent une entrée à chaque partie de cette structure  et les combinent de façon récursive.  Différentes variantes ont été proposées pour certaines tâches du TALN, dont les plus citées dans la littérature sont \cite{Mikolov13} et \cite{bottou13}. Nous renvoyons le lecteur intéressé à l'article de \cite{Bengio13} qui donne un compte-rendu très détaillé de ces différentes avancées.

Il est à noter que ces approches permettent aussi de revenir sur un autre problème de l'indexation de type sac de mots fondée sur des termes simples. Avec ce type d'indexation, l'appariement de plusieurs termes ne différencie pas le cas où ces termes sont en fait toujours liés (comme ceux figurant dans des expressions figées) du cas où ces termes apparaissent peu souvent ensemble. 

\subsection{Synthèse}

Nous nous situons dans le cadre de la recherche d'information au sein de microblogs. La nature de ces données, présentée dans la section~\ref{ss:RImicroblogs}, constitue un défi pour les méthodes de RI classiques : les documents sont trop courts pour pouvoir fournir un contexte propice à une représentation fidèle et sémantiquement fondée. Pour lever ce verrou, il est utile d'aller au delà des simples représentations en sac de mots, en favorisant les représentations distribuées. Ces dernières permettent notamment d'apporter un contexte aux termes qui n'est pas présent dans les documents. 

De très nombreuses méthodes existent pour intégrer ces représentations dans la RI, nous explorons dans cet article les méthodes basées sur l'expansion de requête exploitant des représentations distribuées.

\section{Expansion de requêtes}
\label{sec:exprequ}

Notre approche consiste tout d'abord à fixer un cadre général dans lequel nous allons étudier l'impact des corpus d'apprentissage pour la recherche de microblogs. Comme nous l'avons vu précédemment, les plongements de mots peuvent être utilisés pour étendre les documents ou les requêtes. L'expansion des documents par de tels outils pose clairement la question de l'indépendance des termes, qui dès lors n'est plus liée à la sémantique des documents originaux, mais à un processus de construction. 
Pour se préserver de tels problèmes potentiels, nous utilisons l'expansion de requêtes comme cadre de comparaison.

Une fois le choix de l'utilisation d'expansion de requêtes avec les plongements fait, la question suivante est de déterminer de quelle manière utiliser ces plongements lors de l'expansion de requêtes. Comme nous l'avons vu plus haut, il est possible d'utiliser des approches relativement simples, mais efficaces, pour réaliser cette expansion, en particulier les approches développées par \cite{AlmasriBC16, RoyPMG16}, que nous présentons selon un cadre uniforme ici.

Nous commençons par décrire le résultat de l'apprentissage d'un plongement de mots comme une fonction $pl_p$, de paramètres $p$ définie par:\\
\begin{align}
\label{eq:plong}
pl_p  :  \V & \to  \R^n \\
 w & \mapsto  pl_p(w), \nonumber
\end{align}

où $\V$ est le vocabulaire pour lequel il existe une représentation par $pl$. Ce vocabulaire dépend du corpus d'apprentissage. L'expansion d'une requête $Q$ est donc réalisée par la recherche de termes proches dans le vocabulaire $\V$.

Dans la suite, comme nous fixons a priori tous les paramètres d'apprentissage et que nous nous focalisons sur le corpus d'apprentissage, le paramètre $p$ sera limité au corpus de documents sur lequel les plongements sont appris. 

\subsection{Expansion de requête mono-terme}

Pour une requête ne contenant qu'un seul terme, $Q=\{q\}$, on étend la requête en optimisant une sélection de k-termes du vocabulaire suivant la similarité calculée sur l'espace de plongement. En supposant une fonction de calcul sur tous les $k$-sous-ensembles\footnote{Un k-sous-ensemble d'un ensemble E est un sous-ensemble de E de cardinalité k.} du vocabulaire $V$ dans lequel on cherche les termes d'expansion. Plus précisément, on définit les termes de l'expansion $Exp\_terme$ par :

\begin{equation}
Exp\_termes(q) = \operatorname*{arg\,max}_{S \subset \mbox{ k-sous-ensembles de } V\setminus{q_i}} f(q,S)
\end{equation}
Dans un cas simple, par exemple si $f(q, S) = \sum_{t \in S} sim(q, t)$,
la solution optimale à ce problème peut être calculée trivialement en sélectionnant les k plus proches voisins, notés $NN_k$ t de q suivant la fonction de similarité $sim$, ce qui se réécrit en :

\begin{equation}
Exp\_termes(q) = NN_k(q)
\end{equation}
Certains éléments doivent être détaillés plus précisément pour être totalement décrits. Lorsque l'on utilise un plongement de mots $pl$, l'expression exacte qui opère sur les vecteurs renvoie les k plus proches vecteurs dans l'espace de plongement, suivant une fonction de similarité $sim$ :

\begin{equation}
Exp\_termes\_plongement(q, pl_p, sim) = NN_k^{sim}(pl_p(q))
\end{equation}

Jusqu'à présent, nous avons défini les termes qui seront utilisés lors de l'expansion de requêtes. Comme une requête peut intégrer des éléments d'importance des termes, il semble pertinent de s'en servir pour qualifier les termes étendus.
 Deux éléments supplémentaires viennent s'ajouter à cette étape : i) la pondération de chacun des termes ajoutés à la requête, ii) l'importance globale de l'expansion. Ces deux éléments sont décrits dans la suite.

L'importance d'un terme d'expansion pour une requête mono-terme peut reposer sur l'importance du lien (quelconque) entre le terme d'expansion et la requête initiale. Une telle prise en compte, pour un terme $t$, peut reposer sur la fonction de similarité décrite précédemment : $sim(q,t)$. Dans le cas de plongements, cette similarité se base sur le vecteur de plongement, noté $pl$ des termes considérés : $sim(q,t) = Sim(pl_p(q), pl_p(t))$. L'importance globale de l'expansion est elle considérée de manière simple comme un coefficient multiplicateur, $\alpha$, appliqué à la similarité calculée pour chaque terme.

En intégrant tous ces éléments, nous obtenons l'expression suivante pour l'expansion d'une requête mono-terme :
\begin{multline}
EXP(Q,pl_p,sim,\alpha) = \{(t, \alpha * Sim(pl_p(t), pl_p(q))) | \; t \in \V, pl_p(t) \in NN_k(pl_p(q))\} 
\end{multline}

\subsection{Expansion locale de requête multi-termes : terme par terme}

La première approche que nous considérons pour étendre les requêtes  est celle proposée par~\cite{AlmasriBC16} : l'idée est d'utiliser l'hypothèse que les termes de la requête sont indépendants. Dans ce cas, on recherche séparément par les plongements les termes similaires à chaque terme de la requête. Elle se base sur l'utilisation des $k$ plus proches voisins selon une similarité cosinus des vecteurs de termes, qui est équivalente à un produit scalaire des vecteurs normalisés correspondants. L'idée est qu'un terme de l'expansion de requête doit être lié fortement à, au mininum, un terme de la requête. 
De manière plus formelle, nous supposons qu'une requête $Q$ est composée de couples $(q_i, w_{q_i})_{0\le i \le |Q|}$. Dans le cas d'une requête classique, chaque poids $w_{q_i}$ est égal à 1. Dans ce cas, l'expansion $EXP_{loc}(Q,p)$  de la requête initiale $Q$, avec un plongement $pl_p$ de paramètres $p$, s'exprime par l'ensemble des paires suivant :
\begin{multline}
EXP_{loc}(Q,p) = \bigcup_{q_i \in Q.termes} \{(t, \alpha_{loc} * Sim(pl_p(t), pl_p(q_i))) | \\ \; t \in \V, pl_p(t) \in NN_k(pl_p(q_i)) \}
\end{multline}

Les paramètres de l'expression ci-dessus sont les suivants : $\alpha_{loc}$ dénote le poids du terme dans l'expansion (classiquement moins important que les poids des termes initiaux de la requête) ; $k$ dénote le nombre de plus proches voisins sélectionnés ; la similarité $Sim(.,.)$ mesure la proximité entre les vecteurs de deux termes dans l'espace de plongements (une similarité classique~\cite{mikolovchen2013} est le $cosinus$ des angles des deux vecteurs). Dans l'expression ci-dessus, la notation $Q.termes$ désigne l'ensemble des mots de la requête $\{ q | (w, w_q) \in Q\}$.

\subsection{Expansion globale de requête multi-termes}

La seconde approche étudiée pour étendre les requêtes composées de plusieurs termes repose sur la recherche de termes similaires à toute la requête, elle est similaire à celle proposée par~\cite{Almasri2017} et proche de~\cite{RoyPMG16}. Dans ce cas, l'expression $EXP_{glob}(Q,p)$ s'exprime de la manière suivante :

\begin{multline}
EXP_{glob}(Q,p) = \left\{\left(t, \alpha_{glob} * Sim\left(pl_p(t), \sum_{q_i \in Q.termes} pl_p(q_i)\right)\right) \middle| \right.\\ \;\left. t \in V, pl_p(t) \in NN_k\left(\sum_{q_i \in Q.termes} pl_p(q_i)\right) \right\}
\end{multline}


Dans l'expression ci-dessus : $\alpha_{glob}$ dénote le poids du terme dans l'expansion. On constate qu'un terme ne peut apparaître qu'une fois dans l'expansion. Dans cette configuration, on cherche des termes en relation avec tous les termes de la requête (en fait le vecteur "moyen" de la requête comme on utilise des similarités cosinus). 

\subsection{Requête étendue}

Une fois l'une ou l'autre des expansions effectuées, la requête étendue finale $EXP(Q,p)$ est une fusion de la requête initiale $Q$ et de son expansion $EXP_{x}(Q,p)$ avec $x=loc$ ou $x=glob$. 

Cette fusion a comme objectif d'intégrer de nouveaux termes, mais peut également renforcer un terme existant dans la requête (car il est proche d'autres termes de cette requête). De plus, si un même terme étend plusieurs termes de la requête initiale, alors il va être renforcé (par ajout de chacune des pondérations par terme de la requête).

L'expression finale de $EXP(Q,p)$ est la suivante :\\
\begin{multline}
EXP(Q,p)=\left\{\left(t, \sum_{(t,b) \in Q} b + \sum_{(t,c) \in EXP_x(Q,p)}c\right) \middle|\right. \\ t \in Q.termes \cup EXP_x(Q,p).termes\Bigg\} \nonumber
\end{multline}

Comme nous l'avons dit précédemment, si tous les paramètres d'apprentissage sont fixés, le paramètre $p$ qui va varier dans nos expérimentations se limite au corpus d'apprentissage des plongements.

\section{Expérimentations}
\label{sec:expes}
L'objectif de cette étude est de comparer différents corpus d'apprentissage pour les plongements de mots et leur utilisation pour étendre les requêtes dans le cadre de la recherche d'information. 
Pour cela, nous nous basons sur des systèmes de RI de l'état de l'art, et des méthodes d'expansion de requête présentées dans la section précédente. 
Notre objectif étant de comparer l'apport du corpus d'apprentissage sur les résultats obtenus par l'expansion de requête exploitant les plongements de mots, nous envisageons plusieurs corpus d'apprentissage offrant des caractéristiques différentes : la nature des documents qui les composent ainsi que le domaine dont traitent ces documents. 
ion{Corpus de test} de requê

Nous avons testé nos propositions sur le corpus CLEF CMC 2016 sur la recherche de tweets intitulée ``Timeline Illustration'' (sous-tâche 3,~\cite{GoeuriotMMMS16}). Le corpus complet de CMC contient plus de 50 millions de tweets (contenant le mot "festival" et d'autres termes spécifiques aux festivals considérés). Le sous-corpus que nous utilisons pour la tâche 3 porte sur deux festival musicaux : ``Les vieilles charrues" et les ``Transmusicales", il contient 244 000 tweets.

53 requêtes sont évaluées dans cette sous-tâche. Ces requêtes sont des événements d'un jour entier de chaque festival (par exemple un concert en particulier). Par exemple, la requête $1$ porte sur le concert de ``Khun Narin's Electric'' durant les Transmusicales, le 4 décembre 2015 de 14h à 16h30 :
\begin{verbatim}
<topic>
<id>1</id>
<title>Khun Narin's Electric</title>
<festival>Transmusicales</festival>
<begindate>04/12/15-14:00</begindate>
<enddate>04/12/15-16:30</enddate>
</topic>
\end{verbatim}

Les pertinences ont été établies manuellement sur le {\it pool}.
Les mesures d'évaluations sont classiques : précision au rang 5, 10 et 30 documents, Mean Average Precision (MAP), Mean Reciprocal Rank (MRR). On rappelle que la précision au rang $k$ (p$@k$) pour une requête $q$ est définie comme~:
\[
p@k(q)=\frac{1}{k}\sum_{rg=1}^k R_{d_{rg},q},
\]
où $R_{d_{rg},q}\in\{0,1\}$ est le jugement de pertinence du document de rang $rg$ retourné par le système pour la requête $q$. 

Dans le cas où on dispose d'un ensemble de requêtes $Q = \{q_1, ..., q_{|Q|}\}$, on peut étendre l'évaluation précédente en calculant la moyenne des précisions moyennes (ou MAP) sur l'ensemble des requêtes~:
\[
MAP=\frac{1}{|Q|}\sum_{j=1}^{|Q|} \frac{1}{n_+^{q_j}}\sum_{k=1}^N R_{d_k,q_j}\times p@k(q_j),
\]
Où $n_+^{q_j}$ est le nombre de documents pertinents pour la requête $q_j$. Finalement, la moyenne des rangs réciproques est défini par~:
\[
MRR=\frac{1}{|Q|}\sum_{j=1}^{|Q|} \frac{1}{rang_j},
\]
Où $rang_j$ est le rang du premier document pertinent pour la requête $q_j$. On signale cependant que dans notre cas, les apports les plus notables de notre étude sont obtenus sur les valeurs de précision à 5 documents, nous insisterons donc davantage sur ces valeurs dans la suite.\\

\subsection{Système de recherche d'information}

Pour toutes les expérimentations reportées ici, nous avons utilisé le système de recherche d'information Terrier (V 4.0)~\cite{macdonald2012puppy}. Nous nous sommes basés sur le modèle BM25~\cite{robertson1976relevance} avec les paramètres par défaut proposés par ce système $(k1=1,2; b=0,75; k3=8)$. Dans ce cadre, la pondération des termes de la requête est prise en compte durant le calcul de correspondance par le système. Comme notre corpus est composé de documents en français, nous avons par ailleurs utilisé les prétraitements classiques : antidictionnaire français et troncature de Porter\footnote{Code : \url{http://snowball.tartarus.org/algorithms/french/stemmer.html}}~\cite{van1980new} pour la langue française, proposés par Terrier.

\subsection{Corpus d'entraînement}

Rappelons que notre objectif est de déterminer le comportement du système de recherche de tweets, utilisant de l'expansion de requêtes à base de plongements de mots appris sur différents corpus. Nous avons pour cela choisi d'utiliser quatre corpus d'apprentissage différents :

\begin{enumerate}
    \item[{\bf WF :}] Wikipedia en français - Ce corpus d'apprentissage est contient plus de $1 000 000$ microblogs, issus du {\it dump} wikipedia 2017. Les caractéristiques de ce corpus sont les suivantes : il couvre de nombreux sujets, et les pages peuvent également être assez variées en terme de contenu. L'utilisation de ce corpus permet de déterminer dans quelle mesure un corpus d'apprentissage général à forte couverture thématique (i.e., dont la couverture thématique dépasse les sujets indexés par le SRI) composé de documents de types très différents des tweets, est intéressant à utiliser ;
    \item [{\bf WMF :}] Wikipedia Musical en français - Ce corpus est un sous-ensemble du premier. Il est constitué d'une liste d'environ $55 000$ artistes pour lesquels nous avons extrait de Wikipedia en français les pages, générant un total d'environ $45 000$ articles Wikipedia. Les articles sont donc spécifiques à la musique, mais les pages peuvent être assez disparates. Ce corpus d'apprentissage est donc focalisé sur les sujets similaires au corpus de recherche, mais le type de document (des pages webs encyclopédiques) est très différent des tweets ;
    \item[{\bf TF :}] Tweets généraux en français - $50$ Millons de tweets de sous-tâche 2 de la campagne d'évaluation CLEF $2016$ CMC. Ces tweets couvrent de nombreux sujets. Ici, le choix est donc pris de favoriser la cohérence des types de données (i.e., les tweets) au détriment des sujets ;
    \item[{\bf TMF :}] Tweets musicaux - Les tweets de la tâche 3 de la campagne d'évaluation  CLEF $2016$. Ce corpus contient des tweets de 2 festivals de musique, pour un total de $244 000$ tweets. Pour ce dernier corpus d'apprentissage, on se focalise sur des documents qui sont similaires à ceux ciblés par le SRI (c'est le même corpus), les sujets de ces documents sont donc également focalisés sur ceux du SRI.
\end{enumerate}

Le tableau \ref{tab:train-corpus} résume les informations sur ces corpus. Nous y indiquons en particulier dans quelle mesure les documents des corpus d'entraînement couvrent les mêmes sujets que ceux du corpus de RI, dans quelle mesure les documents d'un corpus d'apprentissage sont similaires à ceux du corpus de RI, et leurs quantités. Ces quantités sont en nombres de documents. Ce tableau nous permet de constater que nous avons une très grande disparité sur la taille de ces corpus en terme de nombre de documents. On peut également signaler que la taille des documents eux-mêmes varie grandement : au moment de l'acquisition de la collection de tweets, ils étaient limités à 140 caractères (donc moins de 10 mots), alors qu'une page Wikipedia, selon les recommandations officielles~\cite{wikiArticleSize}), contient entre 4000 et 7000 mots.  Cette disparité va donc se refléter également sur la taille des vocabulaires (cf. loi empirique de Heaps~\cite{wikiHeapsLaw}) et lignes sans prétraitement, $\emptyset$, de la table ~\ref{tab:voc-corpus}). Quand on sait que les plongements de mots sont classiquement appris sur de grands corpus, on peut se poser la question de savoir si ceux appris sur les tweets TMF seront satisfaisants : est-ce que le fait de se focaliser à la fois sur les sujets et les types de documents contrebalancent la faible quantité de données d'apprentissage ?

\begin{table}
\label{tab:train-corpus}
\caption{Caractéristiques des corpus d'entraînement par rapport au corpus de tweets musicaux du SRI.}
\begin{tabular}{|c|c|c|c|}
    \hline
         \text{Corpus} & \text{Adéquation des sujets} & \text{Adaptation des types  de documents} & \text{\#docs} \\
         \hline
         \hline
         WF & $-$ & $-$ & 1 M \\
         \hline
         WMF & $+$ & $-$ & $45000$\\
         \hline
         TF & $-$ & $+$ & $50$ M\\
         \hline
         TMF & $+$ & $+$ & $245000$\\
         \hline
    \end{tabular}
\end{table}

\subsection{Paramètres étudiés}

\subsubsection{Prétraitement des corpus}

Nous avons choisi d'utiliser les corpus de deux manières différentes : nous les utilisons d'un côté tels quels (sans prétraitement), et de l'autre en les prétraitant suivant une approche classique de RI par application d'un anti-dictionnaire et de la troncature de Porter (en utilisant le même anti-dictionnaire et le même code de troncature que le système de recherche d'information). L'idée que nous voulons explorer ici est de savoir si le fait de rapprocher les données d'apprentissage et du SRI a priori par ces prétraitements a un impact positif sur les résultats. Dans le cas d'une réponse positive, ceci uniformiserait les traitements sur les composantes apprentissage et recherche d'information.

Nous présentons en table~\ref{tab:voc-corpus} les informations liées à la taille des vocabulaires de chacun des quatre corpus, suivant ou non l'utilisation des prétraitements. Nous constatons que l'application de prétraitements pour les pages wikipedia amène à des réductions du nombre de termes de 21\% pour WF et 14\% pour WMF, et sur les tweets de 10\% pour TF et 24\% pour TMF. Ceci souligne que les corpus considérés (pages wikipedia et tweets) ne se comportent pas de la même manière en fonction des prétraitements. 

Dans la suite, nous dénotons les variantes des corpus d'entraînements en indiçant les noms des corpus par $\checkmark$ ou $\emptyset$, suivant que nous y appliquons des prétraitements ou non.

\begin{table}
\label{tab:voc-corpus}
\caption{Variantes des corpus d'entraînements.}
\begin{tabular}{|c|c|c|}

    \hline
         \text{Corpus} & \text{Prétraitement} & \text{\#termes} \\
\hline
\hline
WF & $\emptyset$ & $759\;488$ \\
\hline
WF & $\checkmark$ & $596\;540$ \\
        \hline
\hline
WMF & $\emptyset$ & $63\;106$ \\
\hline
WMF & $\checkmark$ & $43\;786$ \\
        \hline
\hline
TF & $\emptyset$ & $1\;148\;947$ \\
\hline
TF & $\checkmark$ & $1\;028\;488$ \\
        \hline  
        \hline
TMF & $\emptyset$ & $21\;920$ \\
\hline
TMF & $\checkmark$ & $16\;705$ \\
        \hline  
\end{tabular}
\end{table}

\subsubsection{Apprentissage des plongements}

Nous avons utilisé l'outil {\it word2vec}\footnote{https://github.com/dav/word2vec} pour apprendre les plongements sur chacune des variantes de corpus d'apprentissage considérées. Nous avons utilisé comme architecture le modèle CBOW de ce système. Ce choix a été fait car nous avons privilégié la rapidité d'apprentissage des plongements en nous basant sur les tavaux de Mikolov et Chen~\cite{mikolovchen2013}. Les autres hyperparamètres de l'apprentissage que nous avons utilisés sont, à défaut d'information additionnelle, ceux par défaut préconisés par ce logiciel. Ils sont présentés dans le tableau~\ref{tab:hyper-word2vec}. Comme nous utilisons une fenêtre de 8 mots, nous sommes certains que tous les mots d'un tweet seront considérés ensemble.

\begin{table}
\label{tab:hyper-word2vec}
\caption{Valeurs des hyperparamètres utilisés par word2vec.}
\begin{tabular}{|c|c|}

    \hline
         \text{Hyperparamètre} & \text{Valeur}  \\
\hline
\hline
\text{Dimension de l'espace de plongement} & $200$ \\
\hline
\text{Fenêtre} & $8$ \\
        \hline
\text{Nombre d'itérations} & $15$ \\
        \hline
\end{tabular}
\end{table}

Il est à noter que l'utilisation de sous-echantillonnage par word2vec, qui vise à ôter des instances de termes très courants, joue un rôle assez similaire à l'anti-dictionnaire des prétraitements décrits plus haut. Cependant, les prétraitements que nous utilisons éliminent toutes les occurrences des termes de l'antidictionnaire, ce que ne fait pas le sous-échantillonnage de word2vec.

\subsubsection{Paramètres d'expansion}

Nous indiquons maintenant l'ensemble des variantes des paramètres d'expansion que nous avons utilisés dans nos expérimentations. Elles couvrent les méthodes d'expansion, le nombre de termes ajoutés et l'importance de ces termes :
\begin{itemize}
    \item les deux variantes du calcul des termes d'expansion : expansion globale et expansion locale comme décrits en section~\ref{sec:exprequ}. On répond ici à la question de savoir si considérer chaque terme de la requête de manière indépendante est préférable à considérer la requête d'un seul bloc ;
    \item le nombre de termes d'expansion, dans l'ensemble [1, 5]. Ce paramètre correspond au $k$ dans la section~\ref{sec:exprequ}. La question ici est alors de savoir s'il vaut mieux ajouter peu ou beaucoup de termes ;
    \item le poids assigné aux termes étendus. Comme nous l'avons dit précédemment, nous utilisons dans cet article une pondération fixe pour chaque terme étendu. Cette pondération, de manière classique en recherche d'information (cf. Rocchio par exemple), est habituellement moins grande que celle des termes initiaux de la requête, car ces extensions sont calculées et donc moins fiables. Nous avons étudié ici des pondérations dans l'ensemble \{0,1 , 0,2, ..., 0,8 , 0,9\}, pour chacun des paramètres $\alpha_{loc}$ et $\alpha_{glob}$ afin d'en de-éteriner les valeurs les meilleures.
\end{itemize}

\subsection{Résultats}

Cette partie présente les résultats obtenus. Nous avons  choisi quatre éléments prépondérants comme axes d'analyse principaux :
\begin{itemize}
    \item L'impact des prétraitements et des ensembles d'apprentissage pour les plongements, et plus particulièrement l'application d'outils classique de RI : anti-dictionnaire et troncature de mots ;
    \item L'impact des paramètres de pondération des expansions de requêtes pour la recherche d'information, afin de déterminer l'importance à donner aux termes ajoutés à la requête ;
    \item L'impact de la taille des expansions définies, afin de déterminer dans quelle mesure le nombre de termes utilisés pour étendre les requêtes impacte les résultats sur système recherche de tweets.
\end{itemize}

Afin de fournir une base de comparaison, nous commençons tout d'abord par présenter les évaluations des résultats obtenus par le système sans expansion de requête. Elles sont fournies dans le tableau~\ref{tab:res-noexp}.

\begin{table}
\label{tab:res-noexp}
\caption{Les résultats des 5 mesures d'évaluation, sans expansion de requêtes.}
$\begin{array}{|c|c|c|c|c|}
    \hline
         \text{MAP} & \text{MRR} & \text{p@5} & \text{p@10}  & \text{p@30} \\
\hline
\hline
0.0062 & 0,4499 & 0,2718 & 0,2385 & 0,2094  \\
\hline
\end{array}$
\end{table}

Parmi les cinq mesures listées, nous nous concentrons dans la suite sur les résultats de précision à 5 documents (p@5), ce qui permet de mesurer les différences de qualité sur les premiers résultats fournis en réponse, qui sont les plus importants dans les réponses.

\subsubsection{Impact des prétraitements et des corpus d'apprentissage}

Nous décrivons dans les tableaux~\ref{tab:impact-pre-loc} et~\ref{tab:impact-pre-glob} l'impact des prétraitements du corpus d'entraînement (i.e., anti-dictionnaire + troncature) sur les expansions des requêtes. 

Nous nous concentrons sur les paramètres $\alpha_{loc}=0,3$ et $\alpha_{glob}=0,3$ (cf. partie~\ref{sec:impasct_pond}), et nous regardons les résultats en faisant varier le paramètre $k$ de 1 à 5, tous les autres paramètres étant fixés. Dans ces tableaux, nous indiquons les meilleurs résultats par nombre de termes d'expansion (i.e., par colonne) en gras, et nous indiquons en souligné les meilleurs résultats par corpus considéré (i.e., par ligne).
\footnote{Des tests de Student bilatéraux avec un seuil de significativité de 0.05 ont été utilisés pour comparer les résultats de P@5 obtenus avec le résultat sans expansion de requête, et la meilleure de  configuration testée sur la même ligne (souligné). Aucun de ces tests ne s'est avéré concluant.}

\begin{table}
\label{tab:impact-pre-loc}
\caption{Les prétraitements par corpus d'apprentissage, avec $k \in [1, 5]$ pour $EXP_{loc}$ \protect\footnotemark[6]}
$\begin{array}{|c|c|c|c|c|c|}
    \hline
         \text{Corpus} & \text{p@5 | k=1} & \text{p@5 | k=2}  & \text{p@5 | k=3} & \text{p@5 | k=4} & \text{p@5 | k=5}\\
\hline
\hline
WF_\emptyset & \underline{0,2895} & \underline{0,2895} & \underline{0,2895} & \underline{0,2895} & \underline{0,2895} \\
\hline
WF_\checkmark & 0,2842 & \underline{0,2947} & \underline{0,2947} & \underline{0,2947} & \underline{0,2947} \\
\hline
WMF_\emptyset & \underline{0,2789} & \underline{0,2789} & \underline{0,2789} & 0,2737 & 0,2737  \\
\hline
WMF_\checkmark & 0,2789 & 0,2789 & \underline{0,2842} & \underline{0,2842} & \underline{0,2842}  \\
\hline
TF_\emptyset & \underline{0,2737} & \underline{0,2737} & \underline{0,2737} & \underline{0,2737} & \underline{0,2737}\\
\hline
TF_\checkmark & 0,2789 & 0,2895 & 0,3105 & 0,3211 & \underline{0,3263} \\
\hline
TMF_\emptyset & {\bf 0,2947} & {\bf 0,3158} & {\bf 0,3158} & {\bf 0,3316} & \underline{{\bf 0,3421}}  \\
\hline
TMF_\checkmark & 0,2895 & 0.2947 & 0,3105 & 0,3105 & \underline{0,3211}  \\
\hline
\end{array}$
\end{table}

\begin{table}
\label{tab:impact-pre-glob}
\caption{Les prétraitements par corpus d'apprentissage, avec $k \in [1, 5]$ pour $EXP_{glob}$\protect\footnotemark[6].}
$\begin{array}{|c|c|c|c|c|c|}
    \hline
         \text{Corpus} & \text{p@5 | k=1} & \text{p@5 | k=2}  & \text{p@5 | k=3} & \text{p@5 | k=4} & \text{p@5 | k=5}\\
\hline
\hline
WF_\emptyset & \underline{0,2842} & \underline{0,2842} & \underline{0,2842} & \underline{0,2842} & 0,2769 \\
\hline
WF_\checkmark & 0,2789 & \underline{0,2842} & \underline{0,2842} & \underline{0,2842} & \underline{0,2842} \\
\hline
WMF_\emptyset & 0,2737 & 0,2737 & 0,2737 & 0,2737 & \underline{0,2789}  \\
\hline
WMF_\checkmark & \underline{0,2737} & \underline{0,2737} & \underline{0,2737} & \underline{0,2737} & \underline{0,2737}  \\
\hline
TF_\emptyset & \underline{0,2737} & \underline{0,2737} & \underline{0,2737} & \underline{0,2737} & \underline{0,2737} \\
\hline
TF_\checkmark & 0,2737 & 0,2789 & \underline{0,3000} & \underline{0,3000} & 0,2974 \\
\hline
TMF_\emptyset & {\bf 0,2895} & {\bf 0,3105} & {\bf 0,3105} & {\bf 0,3105} & \underline{{\bf 0,3158}} \\
\hline
TMF_\checkmark & 0,2842 & 0,2842 & 0,2947 & 0,2895 & \underline{0,3053}  \\
\hline
\end{array}$
\end{table}

Nous constatons dans le tableau~\ref{tab:impact-pre-loc} que, quelle que soit l'expansion considérée (c'est-à-dire dans 40 cas sur 40), les résultats en termes de P@5 sont meilleurs que ceux sans expansion de requête (cf tableau~\ref{tab:res-noexp}). Ce constat est également établi pour les résultats de l'expansion globale du tableau~\ref{tab:impact-pre-glob}, dans 40 cas sur 40. Il en résulte que ces expansions sont utiles.

De manière globale, les résultats par expansion locale du tableau~\ref{tab:impact-pre-loc} donnent des résultats sensiblement supérieurs à ceux avec expansion globale du tableau~\ref{tab:impact-pre-glob}, dans 35 cas sur 40. Ceci peut provenir du fait que les termes des requêtes (qui sont souvent des noms de groupes) ne sont pas forcément liés dans les plongements (sans doute à cause du faible nombre de co-occurrences dans les corpus d'apprentissage), ce qui fait qu'utiliser de manière conjointe tous les termes dans l'approche globale n'est pas la meilleure méthode.

On se rend compte également que, dans le cas d'expansion locale, l'expansion avec 5 termes donne des résultats au moins égaux à l'expansion avec un seul terme dans 7 cas sur 8. L'exception est celle $WMF_{\emptyset}$ avec une faible décroissance de p@5.
Dans le cas de l'expansion globale, l'expansion à 5 terme fournit également de plus mauvais résultats que l'expansion à un terme dans un seul cas : $WF_{\emptyset}$. Nous tirons de ces résultats deux enseignements : a) dans notre cadre il ne faut pas se limiter à une expansion très courte, et b) pour les plongements appris sur des corpus d'apprentissage très généraux il faut limiter ce nombre.

Un autre élément important que nous tirons des résultats des tableaux~\ref{tab:impact-pre-loc} et~\ref{tab:impact-pre-glob} est que, pour les deux types d'expansion, l'utilisation du corpus d'apprentissage qui correspond au corpus de recherche non prétraité $TMF_\emptyset$ permet d'obtenir les meilleurs résultats, de manière constante, pour tous les k entre 1 et 5, et ceci malgré le fait que ce corpus soit relativement petit. L'adéquation entre les corpus d'apprentissage et de recherche semble donc à privilégier à la taille du corpus d'apprentissage.

Avec les expansions locales, on constate que l'utilisation des prétraitements sur les corpus $WF$, $WMF$ et $TF$ donne des résultats aussi bons ou meilleurs que les corpus non-prétraités. Les prétraitements utilisés ne sont donc pas compatibles avec l'apprentissage des plongements : ceci peut s'expliquer par le fait que les plongements sons dédiés à traiter les mots dans leur forme d'apparition, alors qu'une même troncature peut apparaître dans de nombreux contextes qui ne sont alors pas discriminés.

Dans les deux cadres d'expansions utilisés, les calculs de significativité statistique donnent des différences significatives avec l'approche non-étendue uniquement pour le corpus de tweets musicaux non-prétraités $TMF_\emptyset$ pour une expansion de $k=5$. Ceci renforce encore le fait qu'apprendre les plongements sur le même corpus que celui de recherche non-prétraité est dans notre cas la meilleure solution.

\subsubsection{Impact des pondérations $\alpha_{loc}$ et $\alpha_{glob}$}
\label{sec:impasct_pond}

Nous étudions maintenant les résultats obtenus en fonction de la pondération préfixée pour les expansions de requêtes. Pour cela, nous nous concentrons sur les meilleures configurations pour chaque type d'expansion (locale ou globale), c'est-à-dire $TMF_\emptyset$, sans prétraitement. Les résultats présentés dans les tableaux~\ref{tab:impact-alpha-loc} et~\ref{tab:impact-alpha-glob}, présentent l'impact des variations du paramètre $\alpha$ dans l'intervalle de valeurs $[0,1 ; 0,9]$, par pas de $0,1$, et pour $k$ entre 1 et 5.

\begin{table}
\label{tab:impact-alpha-loc}
\caption{L'impact des valeurs $\alpha_{loc} \in [0,1 , 0,9]$ pour $EXP_{loc}$ avec apprentissage $TMF_{\emptyset}$, pour $k \in [1, 5]$\protect\footnotemark[6].}
$\begin{array}{|c|c|c|c|c|c|}
    \hline
         \alpha_{loc} & \text{p@5 | k=1} & \text{p@5 | k=2}  & \text{p@5 | k=3} & \text{p@5 | k=4} & \text{p@5 | k=5}\\
\hline
\hline
0,1 & 0,2789 & 0,3000 & 0,3053 & 0,3053 & 0,3158 \\
\hline
0,2 & 0,2842 & 0,3053 & 0,3105 & 0,3158 & 0,3263 \\
\hline
0,3 & 0,2947 & 0,3158 & 0,3158 & {\bf 0,3316} & {\bf 0,3421}  \\
\hline
0,4 & 0,3105 & {\bf 0,3316} & {\bf 0,3316} & {\bf 0,3316} & {\bf 0,3421}  \\
\hline
0,5 & {\bf 0,3158} & {\bf 0,3316} & {\bf 0,3316} & {\bf 0,3316} & {\bf 0,3421}  \\
\hline
0,6 & {\bf 0,3158} & {\bf 0,3316} & {\bf 0,3316} & {\bf 0,3316} & {\bf 0,3421}  \\
\hline
0,7 & {\bf 0,3158} & 0,3263 & 0,3263 & {\bf 0,3316} & {\bf 0,3421}  \\
\hline
0,8 & {\bf 0,3158} & 0,3263 & 0,3263 & {\bf 0,3316} & {\bf 0,3421}  \\
\hline
0,9 & {\bf 0,3158} & 0,3263 & 0,3263 & {\bf 0,3316} & {\bf 0,3421}  \\
\hline
\end{array}$
\end{table}

\begin{table}
\label{tab:impact-alpha-glob}
\caption{L'impact des valeurs $\alpha_{glob}$ pour $EXP_{glob}$ avec apprentissage $TMF_{\emptyset}$, pour $k \in [1, 5]$\protect\footnotemark[6].}
$\begin{array}{|c|c|c|c|c|c|}
    \hline
         \alpha_{glob} & \text{p@5 | k=1} & \text{p@5 | k=2}  & \text{p@5 | k=3} & \text{p@5 | k=4} & \text{p@5 | k=5}\\
\hline
\hline
0,1 & 0,2789 & 0,3000 & 0,3053 & 0,3053 & 0,3105 \\
\hline
0,2 & 0,2842 & 0,3053 & {\bf 0,3105} & {\bf 0,3105} & {\bf 0,3158} \\
\hline
0,3 & 0,2895 & {\bf 0,3105} & {\bf 0,3105} & {\bf 0,3105} & {\bf 0,3158}   \\
\hline
0,4 & 0,2895 & {\bf 0,3105} & {\bf 0,3105} & {\bf 0,3105} & {\bf 0,3158}   \\
\hline
0,5 & {\bf 0,2947} & {\bf 0,3105} & {\bf 0,3105} & {\bf 0,3105} & {\bf 0,3158}   \\
\hline
0,6 & {\bf 0,2947} & {\bf 0,3105} & {\bf 0,3105} & {\bf 0,3105} & {\bf 0,3158}   \\
\hline
0,7 & {\bf 0,2947} & {\bf 0,3105} & {\bf 0,3105} & {\bf 0,3105} & {\bf 0,3158}   \\
\hline
0,8 & {\bf 0,2947} & {\bf 0,3105} & {\bf 0,3105} & {\bf 0,3105} & {\bf 0,3158}   \\
\hline
0,9 & {\bf 0,2947} & {\bf 0,3105} &  {\bf 0,3105} & {\bf 0,3105} & {\bf 0,3158}   \\
\hline
\end{array}$
\end{table}

On constate des deux tableaux~\ref{tab:impact-alpha-loc} et~\ref{tab:impact-alpha-glob} que, d'une part les expansions avec le $\alpha = 0,1$ donnent quasiment la même qualité de réponses pour les deux approches, et d'autre part, en regardant par colonne, les résultats atteignent un maximum, aussi bien pour $\alpha_{loc}$ ou $\alpha_{glob}$, entre 0.2 et 0.5 , puis forment un plateau à partir de cette valeur. 

En se focalisant sur comportement de l'expansion globale, dans le tableau~\ref{tab:impact-alpha-glob}, ce plateau arrive pour des valeurs de $\alpha_{glob}$ faibles que dans le cas de l'expansion locale, et la différence de résultats entre les valeurs de $\alpha_{glob}$ égales à 0.2 et 0.3 est moins notable. Dans ce cas, la valeur d'importance des termes ajoutés est donc moins cruciale. Ce constat est moins présent pour les expansions locales.

De plus, on remarque que pour les sélections des 2 ou 3 meilleurs termes dans l'expansion locale, le fait d'accorder une grande importance (i.e., pour $\alpha_{loc} \in [0.7 , 0.9]$) aux expansions dégrade la qualité des réponses, comme le montrent la troisième et la quatrième colonnes du tableau~\ref{tab:impact-alpha-loc}.

On remarque également sur les deux tableaux~\ref{tab:impact-alpha-loc} et~\ref{tab:impact-alpha-glob} que les plus grandes différences pour un $\alpha$ fixé sont obtenues quand $\alpha$ est entre 0.4 et 0.5. Ce point est obtenu juste avant les plateaux. C'est donc autour de ces valeurs qu'il faut choisir les $\alpha$ suivant nos expérimentations.

Dans les cas listés ci-dessus, on retrouve le fait que plus les expansions sont grandes (dans la plage fixée entre 1 et 5), meilleur est le résultat, tout comme nous l'avons constaté dans les expérimentations précédentes.

\subsection{Discussion}


Les résultats que nous avons obtenus ici nous montrent qu'il semble préférable d'utiliser un corpus petit mais bien adapté, plutôt qu'un corpus plus grand mais moins spécifique. En effet, comme le corpus de tweets musicaux que nous avons considéré est relativement petit, on aurait pu supposer que l'apprentissage des plongements ne serait pas intéressant. Les résultats que nous obtenons dans le cadre de nos expérimentations montrent le contraire. 

Comme nous nous situons dans le cas spécifique de la prise en compte des plongements pour la recherche d'information "classique", nous devons aussi garder à l'esprit que nous nous reposons toujours sur l'intersection entre les termes de la requête et les termes des documents. Il en résulte que les termes de l'expansion seront d'autant plus intéressants qu'ils indexent les documents. Utiliser le corpus de recherche pour l'apprentissage des plongements possède donc l'avantage indéniable d'utiliser des termes du vocabulaire d'indexation pour les expansions, ce qui n'est pas garanti sinon, surtout dans le cas de corpus de textes spécifiques.

Les propositions faites ici reposent sur une étape relativement simple pour l'inclusion des plongements. Comme l'a montré~\cite{Billerbeck2004}, l'expansion de requête n'est pas efficace dans tous les contextes, ni sur toutes les requêtes. Une adaptation des paramètres d'expansion aux requêtes et tâches pourrait permettre de limiter le bruit et d'améliorer la recherche de microblogs pertinents. D'autres éléments provenant par exemple de pseudo bouclage de pertinence, comme~\cite{QiangFLY15}, pourrait également aider à affiner les termes d'expansion. Les meilleures solutions seront sans doute dans des combinaisons de ces approches.

\section{Conclusion}
\label{sec:conclusion}
Dans cet article, nous avons étudié l'influence du corpus d'apprentissage utilisé pour un plongement de mots sur la recherche de microblogs culturels en utilisant un corpus de la campagne d'évaluation CLEF CMC 2016. Pour cela, nous avons utilisé deux approches relativement simples pour réaliser des expansions de requêtes à partir de ces plongements pour étudier leur impact. L'avantage de cette simplicité était de permettre l'interprétation des résultats sans l'influence de trop nombreux paramètres. L'évaluation des résultats obtenus en terme de précision à 5 documents nous ont permis de déterminer qu'il est préférable d'apprendre les plongements sur le corpus utilisé par le système de recherche d'information, même si celui-ci est petit par rapport aux travaux de l'état de l'art. Une explication de ce phénomène est que, comme les plongements sont appris sur le même corpus que celui du SRI, il y a moins de 
problèmes de mots hors vocabulaire pour la prise en compte des expansions. Un autre constat que nous avons fait est que l'apprentissage des plongements sur les corpus pré-traités comme en RI (par anti-dictionnaire et troncature) n'est pas du tout adapté à notre cadre de recherche de tweets : il faut donc limiter ces traitements à la partie RI.\\

Dans quelle mesure la taille du corpus et la similarité (typologique ou thématique) influent ce constat initial de manière fine devra faire l'objet d'études ultérieures.
A la suite de notre étude, une question qui se pose est de réussir à profiter conjointement de corpus vastes d'un côté, et de corpus spécifiques de l'autre. Une étude future portera sur la comparaison entre : a) réaliser un premier apprentissage sur un corpus géréral comme Wikipedia, puis continuer l'apprentissage sur un second, ou bien b) uniquement concaténer les corpus en un seul puis réaliser un seul apprentissage sur ce corpus final. Une autre direction pour utiliser plusieurs corpus d'apprentissage serait de s'inspirer de fusion tardive, et d'étudier comment se comportent des expansions intégrant soit les expansions de plusieurs plongements provenant de plusieurs corpus d'apprentissage soit en fusionnant les vecteurs de plongements appris de plusieurs corpus comme dans~\cite{ghannay2016}. 

\section*{Remerciements}
Ce travail a été partiellement financé par les projets :  LIG Emergence 2017 {\it Tonel} et LIG Emergence 2018 {\it Arosoir}.

\bibliographystyle{apalike}

\bibliography{biblio}

\end{document}